\documentclass[%
 reprint,
superscriptaddress,
preprintnumbers,
nofootinbib,
 amsmath,amssymb,
 aps,
prd,
]{revtex4-2}

\usepackage{graphicx}
\usepackage{dcolumn}
\usepackage{bm}
\usepackage{aas_macros}
\usepackage{hyperref}

\newcommand{\lob}{\lambda^{\rm ob}}
\newcommand{\ltr}{\lambda^{\rm tr}}
\newcommand{\lsat}{\lambda^{\rm sat}}
\newcommand{\lcen}{\lambda^{\rm cen}}
\newcommand{\Mmin}{M_{\rm min}}
\newcommand{\dprj}{\Delta^{\rm prj}}
\newcommand{\zob}{z^{\rm ob}}
\newcommand{\ztr}{z^{\rm tr}}
\newcommand{\redmapper}{redMaPPer}
\newcommand{\avg}[1]{\langle #1 \rangle}
\newcommand{\Mpc}{\rm{Mpc}}
\newcommand{\hMpc}{h^{-1}\ \Mpc}
\newcommand{\hMsun}{h^{-1}\ M_{\odot}}

\newcommand{\blob}{b_{\lob}}

\defcitealias{costanzi19}{C19}

\begin{document}

\title[Optical cluster selection bias]{A forward analytical model for the optical selection bias on galaxy cluster lensing profiles}

\author{M.~Costanzi}
\affiliation{Dipartimento di Fisica - Sezione di Astronomia, Università di Trieste, Via Tiepolo 11, 34131 Trieste, Italy}
\affiliation{INAF-Osservatorio Astronomico di Trieste, Via G. B. Tiepolo 11, 34143 Trieste, Italy}
\affiliation{IFPU - Institute for Fundamental Physics of the Universe, Via Beirut 2, 34014 Trieste, Italy}
\author{H.-Y.~Wu}
\affiliation{Department of Physics, Southern Methodist University, Dallas, TX 75205, USA}
\author{J.~H.~Esteves}
\affiliation{Department of Physics, Harvard University, 17 Oxford St, Cambridge, MA, USA}
\author{S.~Grandis}
\affiliation{Universität Innsbruck, Institut für Astro- und Teilchenphysik, Technikerstr. 25/8, 6020 Innsbruck, Austria}
\author{C.~To}
\affiliation{Department of Astronomy and Astrophysics, University of Chicago, Chicago IL 60637, USA}
\affiliation{Kavli Institute of Cosmological Physics, University of Chicago, Chicago IL 60637, USA}
\affiliation{NSF-Simons AI Institute for the Sky (SkAI), 172 E. Chestnut St., Chicago IL 60611, USA}
\author{M.~Aguena}
\affiliation{INAF-Osservatorio Astronomico di Trieste, Via G. B. Tiepolo 11, 34143 Trieste, Italy}
\affiliation{IFPU - Institute for Fundamental Physics of the Universe, Via Beirut 2, 34014 Trieste, Italy}

\begin{abstract}
  
Cluster catalogs selected by optical properties are subject to selection biases, primarily arising from unresolved systems along the line of sight. These biases affect key observables for cluster cosmology, such as the lensing signal and clustering statistics. In this work, we present a fully predictive forward analytical model to quantify the impact of optical-selection bias due to projection effects on cluster density profiles. This is achieved by introducing a scale-dependent parametrization of the optical cluster bias, whose small- and large-scale behaviour is set by the amplitude of projection effects, and by expressing the two-halo component of the density profile in terms of the contributions from off-axis halos along the line of sight. As a case study, we consider a DES Y3-like cluster catalog and validate our model against simulated samples. Our model successfully captures the dependence of the two-halo component on richness boosts induced by projections, as well as its evolution with richness and redshift. It also recovers the overall bias in the projected density profile relative to a randomly selected sample with the same mass distribution. The framework presented here provides a consistent methodology for modeling the impact of line-of-sight structures on the observed richness and density profiles of optically selected clusters, directly linking selection biases to the underlying cosmology and survey specifications.

\end{abstract}


\maketitle



\begin{figure*}
\begin{center}
    \includegraphics[width=0.90 \textwidth]{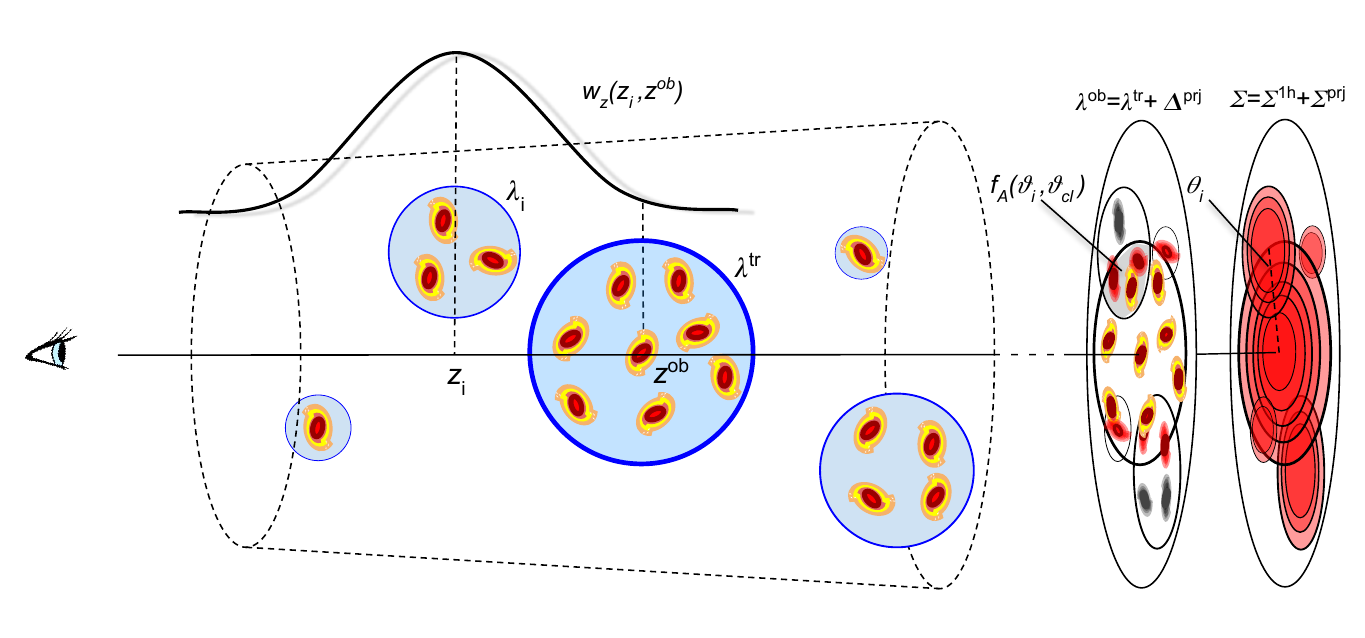}
\end{center}
\caption{Illustration of projection effects on a target cluster with $\ltr$ member galaxies at redshift $\zob$. Each \textit{i}-th halo along the line of sight contributes to the cluster's observed richness with a fraction of its own richness, which depends on the redshift separation and overlapping area: $\rho^{\rm prj}_i = \lambda_i w(z_i,\zob)f_{A,i}$ (see equations \ref{eqn:dprj} and \ref{eqn:weight}). The same systems contribute to the observed projected density profile according to their halo properties and projected angular separation from the target cluster $\theta_i$ (see equation~\ref{eqn:sigprj}).}
\label{fig:sketch}    
\end{figure*}

\section{Introduction}

Galaxy clusters are multi-component systems made of stars, galaxies, hot plasma (the intra-cluster medium), and dark matter. This multi-component nature allows us to detect and study these systems at different wavelengths, such as X-ray, millimeter and optical bands.
In photometric surveys, galaxy clusters are detected as overdensities of galaxies, and the richness of the system -- a weighted sum of the number of putative member galaxies -- is often used as a mass proxy for sample selection. Therefore, systems with their main axis aligned along to the line of sight, more concentrated, and with rich structures in projection have a larger chance to be detected and included in the catalog \citep[e.g.][]{Wu2022,zhangben23,lee25}. In particular, projection effects due to the limited resolution achievable along the line of sight have been shown to bias high the cluster density profile of optically-selected clusters by $10-20\%$ on scales relevant for mass calibration through weak lensing measurements \citep[e.g.][]{Sunayama20,Wu2022,Salcedo2024}. If not properly modeled these effects can lead to biases in the cluster mass estimates \citep[e.g.][]{Nde2025}, which, in turn, translate into systematic biases in the cosmological parameters inferred from optical cluster catalogs \citep[][]{desy1cl}.

Several approaches have been proposed to model or calibrate this selection bias: \citet[][]{Salcedo2024} used a simulation-based approach to emulate stacked cluster lensing profiles and infer observable-mass relation parameters for the DES Y1 cluster sample; \citet[][]{Sunayama23} quantified the large-scale boost of the cluster lensing profile from SDSS data using cluster-galaxy cross-correlation functions \citep[see also][]{Zhou24}. Conversely, in the cosmological analyses of \citet[][]{Sunayama24} and \citet{desy3cl} the authors assumed an analytical template for the selection bias and marginalized over its model parameters.

In this paper, we propose an analytical derivation to estimate the selection bias of optically selected clusters, without free parameters, which depends upon the assumed halo occupation distribution, cosmological model, cluster finder, and survey specifications. We focus on the bias induced by projection effects, which affect scales relevant for weak lensing mass calibration.
In particular, we aim to reproduce the optical selection bias on cluster radial density profiles due to correlated structures as a function of richness and redshift. 
To this end, we first model the clustering bias of a richness-selected sample, and then derive the corresponding projected density profile.
As a case study, we assume an optically selected cluster catalog built with the red sequence Matched-filter Probabilistic Percolation cluster finder algorithm \citep[redMaPPer;][]{Rykoff2014} using Dark Energy Survey\footnote{ https://www.darkenergysurvey.org; DES Year 3 data release: https://des.ncsa.illinois.edu/releases/y3a2/Y3key-cluster.} (DES) Year 3-like photometric data \citep{desy3cl}. Comparing our model predictions to synthetic data, we show the capability of our model to reproduce the scale dependence and amplitude of the bias, as well as its dependence on the strength of projection effects. Despite its simplicity -- e.g. we neglect orientation and concentration bias -- the model provides a useful tool to estimate the impact of projection effects on cluster density profiles on scales relevant for their mass calibration, which can be tuned to the specifics of the cluster finder and survey data.

The paper is organized as follows. In section~\ref{sec:model}, we present the model for the optical cluster bias and the projected density profile. Section~\ref{sec:cos} describes the case study and the construction of the mock cluster catalog. The analysis and interpretation of the results, together with comparisons to synthetic data, are provided in section~\ref{sec:res}. Finally, section~\ref{sec:conc} summarizes our findings and conclusions.

\section{Model}
\label{sec:model}

Our model builds on the observation that the same line-of-sight structures that contribute to the observed richness also affect the projected density profile (see e.g. figure~\ref{fig:sketch}). Consequently, a larger projection-induced boost in richness, $\dprj$, implies a more crowded environment along the line of sight, which in turn enhances the projected density profile. Therefore, if we can predict the mean number of halos projected along the line of sight of a cluster as a function of $\dprj$ and angular separation $\theta$ from its center, we can also predict their contribution to the observed density profile. To this end, we first derive an expression for the optical cluster bias as a function of $\dprj$ and projected angular separation (section~\ref{sec:bias}). This relation then allows us to estimate the mean contribution of correlated structures to the projected density profile by summing the contributions of halos projected at different off-axis angles $\theta$ (section~\ref{sec:sigma_prj}).

The framework developed in this work builds upon the approach introduced by \citet[][hereafter C19]{costanzi19}, in which the observed richness, $\lob$, is treated as a noisy measurement of the intrinsic (true) richness, $\ltr$ -- i.e. the unobservable true number of member galaxies satisfying the survey selection criteria -- which depends on cluster mass.
Specifically, we assume $\lob = \ltr + \Delta^{\rm prj}$, where $\Delta^{\rm prj}$ accounts for the contribution from projected structures along the line of sight \citep[e.g.][]{farahietal16,zuetal17,Wojtak2018}, while $\ltr$ includes any other source of intrinsic scatter.
The stochastic nature of $\lob$ is described by decomposing the probability of observing a system of mass $M$ and redshift $z$ with richness $\lob$ as the convolution:
\begin{equation}
\label{eqn:Plobmz}
 P(\lob | M,z) = \int_0^{\infty} d \ltr \ P(\lob | \ltr,z) P(\ltr | M,z) \, .
\end{equation}
Here $P(\ltr | M,z)$ encodes the intrinsic richness-mass relation, while $P(\lob | \ltr,z)$ accounts for the observational scatter, which depends on the specifics of the survey and cluster-finding algorithm. The latter has been calibrated in the literature either through simulation-based approaches \citep[e.g.][]{costanzi19,Maturi19} or via multi-wavelength observations \citep[e.g.][]{Grandis25}.

Throughout this work, we neglect the cluster photometric redshift uncertainty, and set $\zob$ equal to the true cluster redshift. The latter has been shown to be a subdominant systematic in current cluster cosmology studies \citep[e.g.][]{desy1cl,SPT25,kidscl1000}, but it can be restored in the following equations at the expense of an additional marginalization over the distribution $ P(\zob|\ztr) d \ztr$. The formalism outlined below is general and does not rely on assumptions specific to a particular optical survey. The specific model choices and survey-dependent ingredients adopted for this study are presented in section~\ref{sec:cos}.

\subsection{Optical cluster bias}
\label{sec:bias}
 
Given two tracers of the matter density field, the mean number of tracer "1" neighbors that each tracer "2" has within a volume element $dV$, separated by a distance $r$, is:
\begin{equation}
    dN(r) = \bar{n}_1 \left[ 1 + b_1 \, b_2 \xi(r)\right] dV \, ,
\end{equation}
where $\bar{n}_1$ is the mean number density of tracer "1", $\xi$ is the matter correlation function, and $b_1$ and $b_2$ the biases of the two tracers. For our purposes -- predicting the mean number of halos in projection onto an optical cluster -- $b_1$ is the halo bias and $b_2 = \blob$, the optical cluster bias, which we aim to model as a function of observable cluster properties.

To this end, we start by expressing the noise on the observed richness due to projection effects, $\dprj$, as the weighted sum of the foreground and background cluster galaxies falling within the target cluster area \citepalias[see also Appendix A in][]{costanzi19}. Specifically, for clusters at redshift $\zob$ with observed richness $\lob$, and true richness $\ltr$, the mean $\dprj$ due to projected systems reads:

\begin{multline}
\label{eqn:dprj}
\Delta^{\rm prj}(\lob,\ltr,\zob) = \lambda^{\rm ob}-\lambda^{\rm tr} =2 \pi \int dz \frac{d V}{d z d \Omega}  \int_0^{\pi} d \theta \sin\theta \\  \cdot \int d\lambda \, \rho^{\rm prj}(\lambda,z,\theta|\lob,\zob)  n^{\rm prj}(\lambda,z,\theta|\lob,\ltr,\zob) \, ,
\end{multline}
where $d V /(d z d \Omega)$ is the comoving volume element per unit redshift and solid angle, $n^{\rm prj}(\lambda,z,\theta|\lob,\ltr,\zob)$ is the mean number density of halos at redshift $z$ with true richness $\lambda$, at projected angular distance $\theta$ from the target cluster center, while $\rho^{\rm prj}(\lambda,z,\theta|\lob,\zob)$ is the mean richness contamination contributed by each halo in projection\footnote{To avoid clutter in the notation, we denote the true richness of the target cluster as $\ltr$, while using simply $\lambda$ for the true richness of projected halos.}. This quantity is specific to the cluster finder and survey data under consideration. We factorize the latter term as:
\begin{equation}
\label{eqn:weight}
    \rho^{\rm prj}(\lambda,z,\theta|\lob,\zob) = w_z(z,\zob)f_A(\lambda,z,\theta,\lob,\zob) \lambda \, ,
\end{equation}
where $w_z(z,\zob)$ is the mean fraction of richness that a cluster will absorb from a halo at 
redshift $z$, assuming perfect alignment of the two systems. The second term in equation \ref{eqn:weight} accounts for the geometry of the systems, and corresponds to the mean fraction of galaxies of a halo falling within the target cluster projected area. 

The mean number density of halos in projection is computed according to the cluster-halo correlation function as:
\begin{multline}
    \label{eqn:n}
  n^{\rm prj}(\lambda,z,\theta|\lob,\ltr,\zob)  = \int dM n(M,z) P(\lambda|M,z) \, \cdot\\ \left[ 1 + b(M,z)\blob(\lob,\ltr,\zob,\theta) \xi_{\rm NL}(z,\zob,\theta)\right] \,
\end{multline}
where $n(M,z)$ represents the halo mass function, $P(\lambda|M,z)$ the probability distribution of the true richness at fixed halo mass and redshift, $b(M,z)$ is the halo bias, and $\xi_{\rm NL}(z,\zob,\theta)$ the non-linear matter correlation function at $\zob$ and comoving separation:
$$
\Delta \chi(z,\zob,\theta)=\sqrt{\chi(z)^2+\chi(z^{\rm ob})^2 - 2\chi(z)\chi(z^{\rm ob})\cos \theta } \,.
$$ 
Finally, $\blob$ is the scale-dependent bias for clusters with $\lob$, $\ltr$ and $\zob$ that we seek to predict. To account for exclusion effects, we include the condition
$
 b(M,z) \blob(\lob,\ltr,\zob,\theta) \xi_{\rm NL}(z,\zob,\theta)= -1  
$
for comoving distances $\Delta \chi$ smaller than the cluster radius $R_\lambda$ defined by the detection algorithm.

Since only structures projected within $R_\lambda$ contribute to $\lob$, and since correlated structures are typically distributed along filaments rather than isotropically around the cluster, the selection at fixed $(\lob,\ltr)$ preferentially picks systems with either denser (\mbox{$\lob-\ltr\gg1$}) or emptier ($\lob\simeq\ltr$) lines of sight. As a consequence, the modulus of $\blob$ is expected to decrease with projected distance from the cluster center, eventually reaching a constant value, $\blob^\infty$, beyond $R_\lambda$ \citep[see also][]{Zeng2023,Sunayama23,desy3cl_method}.
We therefore propose the following sigmoid functional form:
\begin{multline}
\label{eqn:b_sel_theta}
\blob(\lob,\ltr,\zob,\theta) = \blob^0(\lob,\ltr,\zob) \, + \\ \frac{\blob^\infty(\lob,\ltr,\zob) - \blob^0(\lob,\ltr,\zob)}{1 + e^{-k(\theta-\theta_0)}} \, ,
\end{multline}
where $k$ controls the steepness of the transition, $\theta_0$ defines its midpoint, while $\blob^0$ and $\blob^\infty$ represent the asymptotic values on small- and large-scale, respectively. The model is motivated by the analysis of the ratio of  projected correlation functions measured from our mock cluster catalog (see section \ref{sec:mocks}) using a richness-selected and a random-selected sample with the same mass distribution (see Appendix \ref{app:bsel} for details). In particular, the mock data are well described by the sigmoid model setting the shape parameters to $k=2.5 / \theta_\lambda$ and $\theta_0 = \theta_\lambda/2$, where the cluster angular size is given by $\theta_\lambda = R_\lambda / D_a(\zob)$.
As for the large-scale bias, $\blob^\infty$, it is well approximated by the relation:
\begin{eqnarray}
\label{eqn:bsel_infty}
    &\blob^\infty(\lob,\ltr,\zob) = b_{\rm eff}(\lob,\zob) \left[ 1 + 0.13 \delta^{\rm prj}\right] \nonumber \\
    &\delta^{\rm prj}(\lob,\ltr,\zob) = \frac{\lob - \ltr}{\dprj_{\rm RND-sel}(\lob,\zob)} -1  
\end{eqnarray}
where $\delta^{\rm prj}$ corresponds to the excess of $\dprj$ compared to randomly selected sample with the same mass distribution (see figure \ref{fig:b_sel_infty}). The effective bias is computed by weighting the halo bias by the corresponding mass distribution: $b_{\rm eff}(\lob,\zob) = \int dM P(M|\lob,\zob) b(M,\zob)$, while the mean contribution to the richness for an unbiased (mass-selected) sample, $\dprj_{\rm RND-sel}$, is derived from equations \ref{eqn:dprj}-\ref{eqn:n}, by replacing $\blob$ with $b_{\rm eff}$.

Using our model for $\blob$, we can invert equation~\ref{eqn:dprj} to obtain an explicit expression for the  small-scale bias term $\blob^0(\lob,\ltr,\zob)$:
\begin{align}
\label{eqn:bsel0}
\blob^0(\lob,\ltr,\zob)
&=
\frac{
(\lob-\ltr)
-
\avg{\Delta^{\rm prj}_{\rm bkg}}
-
\blob^\infty I_1
}{
I_2-I_1
}
\\
\avg{\Delta^{\rm prj}_{\rm bkg}(\lob,\zob)}
&=
\mathcal{P}[1] ,  \tag*{}
\\
I_1(\lob,\zob)
&=
\mathcal{P}\!\left[
b(M,z)\,
\frac{\xi_{\rm NL}(z,\zob,\theta)}
{1+e^{-k(\theta-\theta_0)}}
\right] ,  \tag*{}
\\
I_2(\lob,\zob)
&=
\mathcal{P}\!\left[
b(M,z)\,
\xi_{\rm NL}(z,\zob,\theta)
\right] ,  \tag*{}
\end{align}
where we define, for convenience, the operator $\mathcal{P}[\cdot]$ as the line-of-sight integral over the halo population weighted by the projection kernel $\rho^{\rm prj}$: 
\begin{multline}
\label{eqn:prj_defs}
\mathcal{P}[X]
\equiv
\int dz\,\frac{dV}{dz\,d\Omega}
\int dM\, n(M,z) 
\int d\lambda\, P(\lambda|M,z) \\
\cdot 2\pi \int d\theta\,\sin\theta\;
\rho^{\rm prj}(\lambda,z,\theta|\lob,\zob)\,
X(M,z,\theta|\zob) \, .
\end{multline}

From equation \ref{eqn:bsel0}, it is evident that the bias at small projected distance, $\blob^0$, depends on the richness boost, $\Delta^{\rm prj} = \lob - \ltr$, relative to the expected contributions from the background, $\langle \Delta^{\rm prj}_{\rm bkg} \rangle$, and correlated large-scale structures, $\blob^\infty I_1(\lob,\zob)$.

\begin{figure}
    \includegraphics[width=0.48 \textwidth]{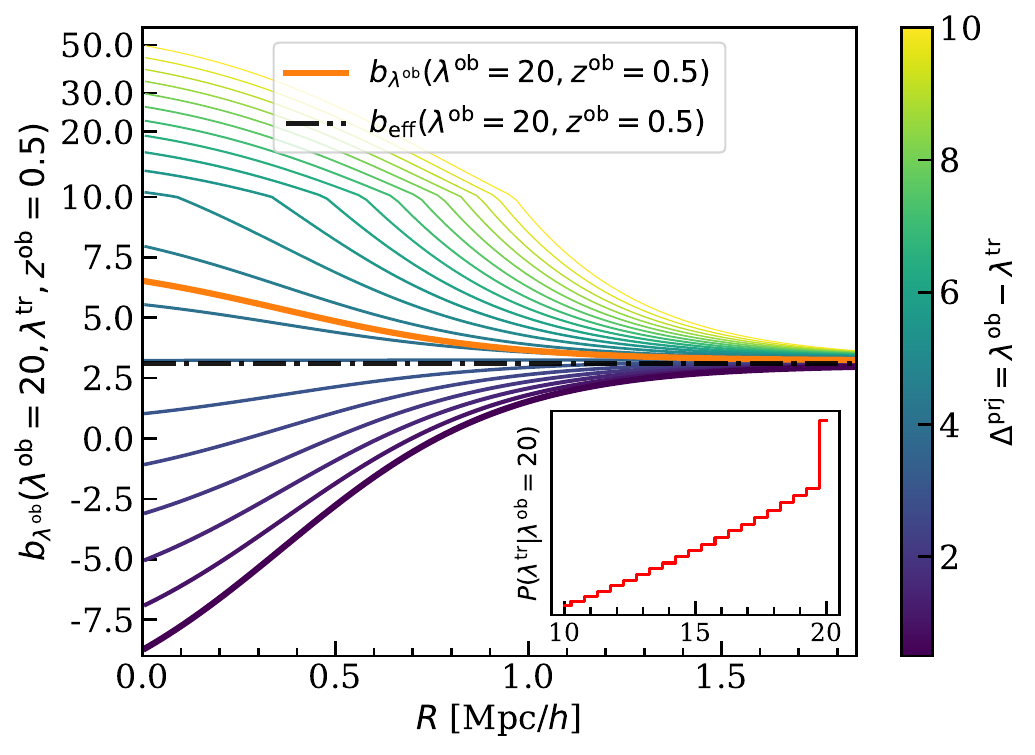}
\caption{Optical cluster bias as a function of radial distance for different $\ltr$ values (color-coded) and fixed $\lob=20$, $\zob=0.5$ (equation \ref{eqn:b_sel_theta}). For ease of interpretation, we convert the projected angular separation to the physical distance $R=\theta D_A(\zob)$. Larger $\Delta^{\rm prj}=\lob-\ltr$ values correspond to higher bias values within the projected cluster radius $R_\lambda$. For $\Delta^{\rm prj}$ below the background contribution ($\sim 2$), the bias becomes negative, corresponding to preferentially under-dense lines of sight selected at fixed $(\lob,\ltr)$. The \textit{orange} line represents the mean optical bias for $\lob=20$ clusters (equation \ref{eqn:b_sel}) computed by weighting $\blob(\lambda^{\rm ob},\lambda^{\rm tr},z,\theta)$ by the distribution $P(\ltr|\lob=20)$ (equation \ref{eqn:P_dprj}) which is displayed in the inset plot with arbitrary normalization. For comparison, the expected bias for a randomly selected sample with the same mass distribution, $b_{\rm eff}$, is shown with a dot-dashed \textit{black} line. Above $\blob=10$ the y-axis is shown on a logarithmic scale to increase the displayed dynamical range.}
\label{fig:bsel}    
\end{figure}

To illustrate the model behavior, we show in figure \ref{fig:bsel} the optical cluster bias expected for $\lob=20$ and $\zob=0.5$ clusters assuming different richness contamination values (color-coded lines). Depending on the amplitude of $\dprj$ compared to the average background contribution, the model predicts either an excess or deficit of correlated structure around the target cluster line of sight.

Finally, the mean bias for clusters observed with $\lob$ can be derived from equation \ref{eqn:b_sel_theta} by marginalizing over the $\ltr$ distribution, namely:
\begin{multline}
\label{eqn:b_sel}
 \blob (\lambda^{\rm ob}, z^{\rm ob},\theta) = \int d\lambda^{\rm tr} \blob(\lambda^{\rm ob},\lambda^{\rm tr},z,\theta) P(\ltr|\lob,\zob) \\ 
  = \frac{1}{n(\lob,\zob)}\int d\lambda^{\rm tr} \blob(\lambda^{\rm ob},\lambda^{\rm tr},z,\theta) P(\lambda^{\rm ob}|\lambda^{\rm tr},z^{\rm ob}) \\ \cdot \int dM P(\ltr|M,\zob) n(M,\zob) \, .
\end{multline} 

The resulting mean bias for $\lob=20$ clusters at $\zob=0.5$ is shown in figure~\ref{fig:bsel} with a \textit{orange} solid line. Compared to a random-selected sample with the same underlying mass distribution $P(M|\lob,\zob)$ (\textit{black} dot-dashed line), the optical cluster bias at $R=0$ is $2.1$ times larger, while on large scales it is boosted by $5\%$.

\subsection{Projected density profile}
\label{sec:sigma_prj}
Following the same line of thought outlined in section \ref{sec:bias} we can construct a model for the mean azimuthally-averaged
projected density profile of an optically selected cluster having $\lob$ and $\zob$.

We start by splitting the mean projected density profile into two contributions \citep[e.g.][]{Oguri2011}:
\begin{equation}
\langle \Sigma (R|\lambda^{\rm ob},z^{\rm ob}) \rangle = \langle \Sigma^{\rm 1h} (R|\lambda^{\rm ob},z^{\rm ob}) \rangle + \langle \Sigma^{\rm prj} (R|\lambda^{\rm ob},z^{\rm ob}) \rangle \, ,
\end{equation}
where the first term represents the contribution of the one-halo term, and can be computed by properly weighting the projected density profile of halos with mass $M$ and redshift $\zob$, $\Sigma(M,\zob)$, with the underlying mass distribution at fixed $\lob$:
\begin{multline}
    \label{eqn:sig1h}
    \langle \Sigma^{\rm 1h} (R|\lambda^{\rm ob},z^{\rm ob}) \rangle = \int dM P(M|\lob,\zob) \Sigma(R|M,\zob) \\ = \frac{1}{n(\lob,\zob)}\int d M P(\lob|M,\zob) n(M,\zob) \Sigma(R|M,\zob)
\end{multline}

The second term -- commonly referred to as the two-halo term -- accounts for the contribution from correlated large-scale structures. In this work, we express it as an integral over the contributions, at projected distance $R$, from halos along the line of sight centered at different angles $\theta$ from the target cluster. Using the cluster bias $\blob$ derived in the previous section to compute the number of halos in projection to a target cluster with ($\lob,\zob$), the 2-halo term reads:
\begin{multline}
    \label{eqn:sigprj}
    \langle \Sigma^{\rm prj} (R|\lambda^{\rm ob},z^{\rm ob}) \rangle = \int d \theta \sin\theta \int d z \frac{d V}{d z d \Omega} 
    \int d M\ n(M,z) \\  \left[1 + b(M,z) \blob (\lambda^{\rm ob}, z^{\rm ob},\theta) \xi_{\rm NL}(z,\zob,\theta)\right] \Sigma_{\rm mis} (R |M,z, \theta,\zob) \, .
\end{multline}

The term $ \Sigma_{\rm mis} (R |M,z, \theta,\zob)$ is the surface mass density contributed at the projected distance $R$ from the target cluster by halos in projection with mass $M$, at redshift $z$, and miscentered by an angle $\theta$ relative to the target cluster center \citep[e.g.][]{Yang2006,Johnston2007}: 
\begin{eqnarray}
\label{eqn:sig_mis}
 \Sigma_{\rm mis} (R |M,z,\theta,\zob) =  \int_0^{2 \pi} \Sigma\left(R_h| M,z \right) \, {\rm d} \varphi, \\
 R_h = \frac{\sqrt{\theta_R^2+\theta^2-2\theta\theta_{R}\cos\varphi}}{D_A(z)} \, ; \quad \theta_R=\frac{R}{D_A(\zob)} \, , \nonumber
\end{eqnarray} 
where the integral over $\varphi$ accounts for the contributions of systems at different azimuthal positions. Equation \ref{eqn:sig_mis} properly accounts for the different redshifts of the target cluster and the halos in projection.

\begin{figure}
    \includegraphics[width=0.48 \textwidth]{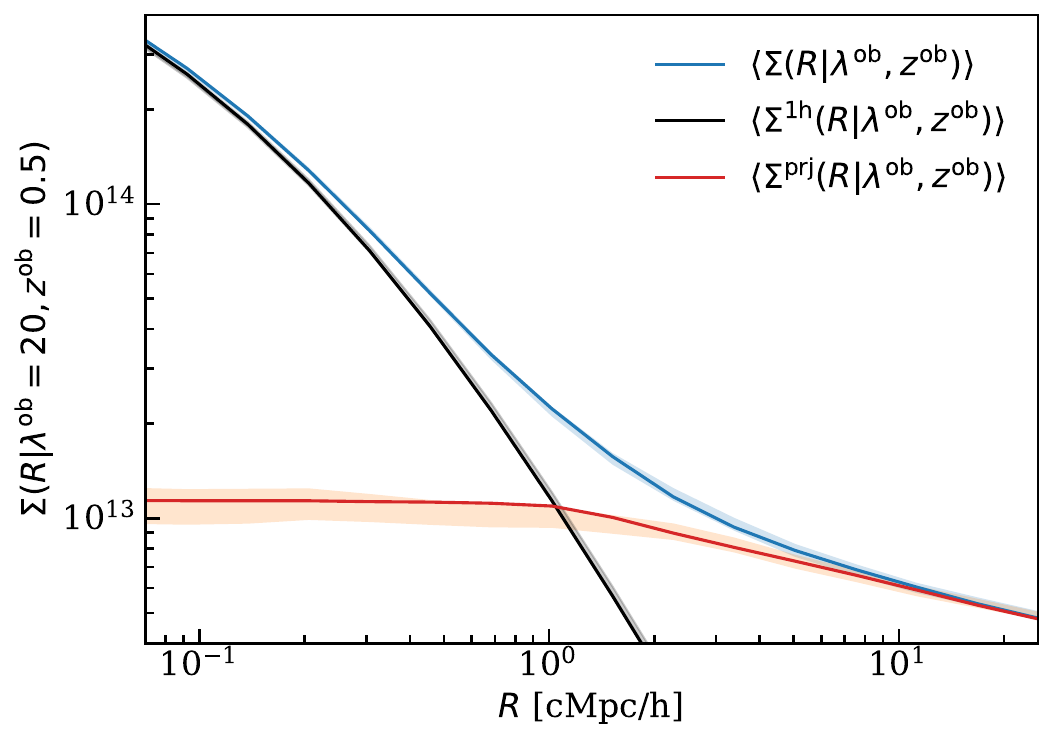}
\caption{Mean projected density profile for $\lob=20$ and $z=0.5$ (\textit{blue} line). The \textit{black} line shows the contribution from the one-halo term (equation \ref{eqn:sig1h}), while the \textit{red} line the contribution from halos in projection (equation \ref{eqn:sigprj}). The contribution of the latter becomes dominant above $R \sim 1 \hMpc$. The \textit{shaded} bands represent the mean profiles retrieved from the mock catalog with their $1\sigma$ uncertainty (see section~\ref{sec:mocks}).}
\label{fig:sig_prof}    
\end{figure}

For illustrative purposes, we show in figure \ref{fig:sig_prof} the resulting mean density profile for clusters with observed richness $\lob=20$ at redshift $z=0.5$,  along with its two components, $\avg{\Sigma^{1h}}$ and $\avg{\Sigma^{\rm prj}}$. 

%

\begin{figure*}
\begin{center}
    \includegraphics[width=0.85 \textwidth]{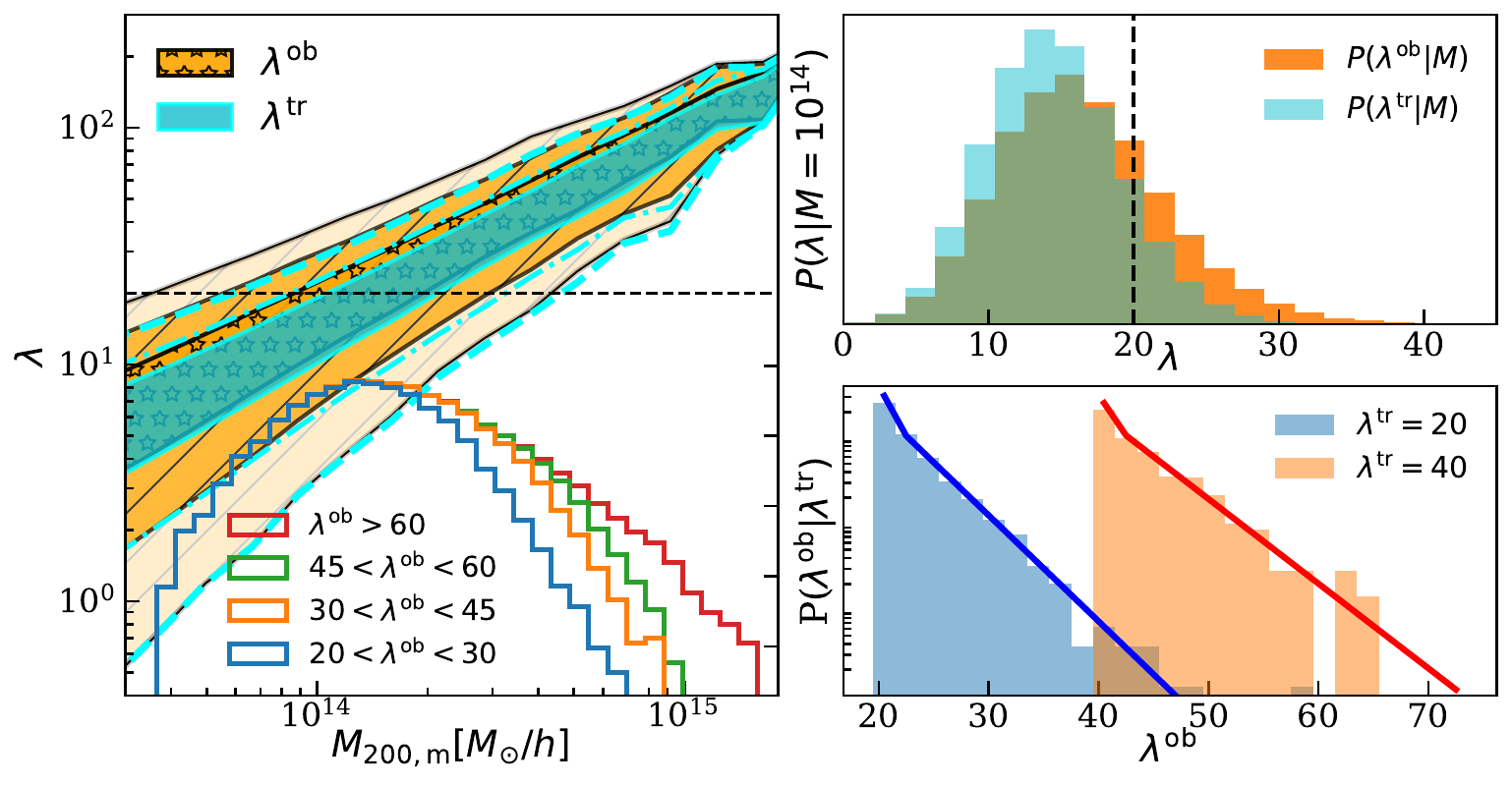}
\end{center}
\caption{Scaling relations of the mock DES Y3 cluster catalog in the redshift range $0.2<z<0.65$: \textit{Left} panel: $\lob$ and $\ltr$-mass relations; the \textit{cyan} and \textit{hatched orange} bands represent the 68, 95 and 97.5 percentile of the $\ltr$ and $\lob$ distributions, respectively. The stacked histograms at the bottom shows the mass distributions of the cosmological sample ($\lob>20$) for the different richness bins adopted in the analysis. \textit{Upper right} panel: Probability distribution functions of $\lob$ (\textit{orange}) and $\ltr$ (\textit{cyan}) for $M=10^{14} \hMsun$ systems.  \textit{Lower right} panel: $P(\lob|\ltr)$ distributions for $\ltr=20$ and $40$. The solid lines correspond to the model predictions derived from equation \ref{eqn:P_dprj}.}
\label{fig:OMR}    
\end{figure*}

\section{Case study: \redmapper\ DES Y3}
\label{sec:cos}

To test our model, we assume a DES Y3-like cluster catalog, built
using the \redmapper\ cluster finder \citep{Rykoff2014}. 
\redmapper\ is a photometric cluster finding algorithm that identifies galaxy clusters as overdensities of red-sequence galaxies and then calculates the richness as the sum of the membership probabilities of all galaxies in the cluster field. The sum extends over all red-sequence galaxies above a fixed luminosity threshold and within an empirically calibrated cluster radius $R_\lambda = (\lob/100)^{0.2} \hMpc$.
Without loss of generality, we make the following model choices: 
\begin{itemize}
    \item We use the Tinker halo mass function and halo bias models \citep{Tinker2008,Tinker2010};
    \item We model the projected density $\Sigma(R|M,z)$ assuming a spherical NFW profile \citep[][]{NFW}.
    \item The intrinsic richness--mass relation is modeled using a conventional halo model parameterization, with $\ltr = \lcen + \lsat$ where $\lcen$ and $\lsat$ are the number of central and satellite galaxies respectively.  $\lcen$ is assumed to be a deterministic function of mass, with $\lcen=1$ for $M\geq \Mmin$ and $\lcen=0$ otherwise. 
$\lsat$ is a random variable with an expectation value
\begin{equation}
\label{eqn:RMR}
\avg{\lsat|M,z} = \left( \frac{M-\Mmin}{ M_1 - \Mmin} \right)^\alpha \left( \frac{1 + z}{1 + z_*} \right)^\epsilon 
\end{equation}
where $M_1$ is the characteristic mass at which a halo of mass $M$ has on average one satellite galaxy, and the pivot redshift is set equal to the mean redshift of the sample $z_*=0.45$.
To allow for super-Poisson halo occupancies, we model $P(\ltr|M,z)$ as the convolution of a Poisson and a Gaussian distribution, where the scatter of the latter is given by $\sigma_\lambda \avg{\lsat|M,z}$. 
\item To mimic the percolation scheme adopted by \redmapper\, $\rho^{\rm prj}(\lambda,z|\ltr,\zob,\theta)$ is set to zero if $\lambda \geq \lob$, that is, only lower-ranked clusters can contribute to the richness of a given object\footnote{We note that a full implementation of the \redmapper\ percolation scheme would also require accounting for the loss of member galaxies to higher-ranked systems. Given the small fraction of clusters affected by this effect \citepalias[see e.g. Fig.~8 of][]{costanzi19}, and its marginal impact on the $P(\lob|\ltr)$ relation, we defer the implementation of this contribution to future work.}. This effectively corresponds to setting the upper limit of the integral over richness in equation \ref{eqn:dprj} to $\lob$. For the same reason, only halos at projected separation $\theta \lesssim 2\theta_\lambda$ can contribute to $\Delta^{\rm prj}$. 
\item Following \citetalias{costanzi19} we model the redshift weight in equation \ref{eqn:weight} as:
 \begin{equation}
 \label{eqn:w_z}
 w_z(z,\zob)= \left \{ \begin{array}{rl}1-\frac{(z-\zob)^2}{\sigma_{z}(z)^2} \; , & |z-\zob|<\sigma_z(z) \\ 0 \; , & \mathrm{otherwise} \end{array} \right .
 \end{equation}
where the width of the parabolic kernel, $\sigma_z$ -- a measure of the photo-z uncertainty of the data -- is calibrated using DES Y3 data following the method detailed in \citetalias{costanzi19} (see their section 3.2).
\item The distribution $P(\lob|\ltr,z)$ is approximated as the sum of a delta function and an exponential component, following the simulation-based prescription of \citetalias{costanzi19} (their equation~5):
\begin{equation}
\label{eqn:P_dprj}
P(\lob | \ltr,z) = (1-f^{\rm prj})\delta_D(\lob-\ltr)+ f^{\rm prj} \tau e^{-\tau (\lob-\ltr)} \, ,
\end{equation}
whose parameters $f^{\rm prj}$ and $\tau$ are computed, as a function of $\ltr$ and $z$, according to equation A8 in \citetalias{costanzi19}.
\item Assuming the cluster galaxies to be uniformly distributed within $R_\lambda$, the geometrical term $f_A(\theta,\ltr,\zob,\lambda,z)$ corresponds to the fraction of overlapping area of the clusters in projection; for the sake of brevity, we report the analytic expression of this term in appendix \ref{app:a}. We explicitly checked using our mock (see section~\ref{sec:mocks}) that assuming a more realistic radial distribution for the member galaxies does not significantly affect our results.
\item We neglect photo-z errors in the cluster redshift estimate and assume that the observed redshift $\zob$ is equal to the true cluster redshift.
\item We exclude miscentering effects from the analysis and set the observed cluster center equal to the true halo center. 
\end{itemize}

\subsection{Synthetic Validation Catalogs}
\label{sec:mocks}
To validate our core analytical framework -- specifically the modeling of the small-scale selection bias (Eq. \ref{eqn:bsel0}) and the projected density profile component (Eq. \ref{eqn:sigprj}) 
we generate a synthetic DES Y3-like cluster catalog according to the model choices outlined in section~\ref{sec:cos}. 
We start from the halo catalog derived from the $N$-body simulation of \citet{DeRose2019BuzzardFlock} based on a flat-$\Lambda$CDM cosmological model with parameters: $\Omega_{\rm m} = 0.286$, $h_0 = 0.7$, $\Omega_{\rm b} = 0.047$, $n_s = 0.96$, and $\sigma_8 = 0.82$. The simulation, which resolves $1400^3$ particles within a cubic volume of $[1050\ h^{-1}\ {\rm Mpc}]^3$, was executed using the L-Gadget code, a modified version of Gadget \citep{Springel2005}. A lightcone, spanning a quarter of the sky over redshift $0.1 < z < 0.9$, was generated from the simulation. Halos were identified using the {\tt Rockstar} halo finder \citep{Behroozi2013} down to $M_{200m}=10^{12.5} M_\odot / h$. For this analysis we consider all halos with $M_{200m}>10^{13} \hMsun$\footnote{All masses are defined with respect to an overdensity of 200 relative to the mean.}.

We assigned an intrinsic richness to each halo according to its mass and redshift by drawing values from the probability distribution $P(\ltr|M,z)$ outlined in section~\ref{sec:cos}, with parameters: $M_{\rm min}=10^{11.4}$, $M_1=10^{12.7}$, $\alpha=0.86$, $\epsilon=0.28$ and $\sigma_\lambda=0.18$\footnote{These parameter values have been derived by fitting the DES Y1 cluster counts, assuming the DES Y1 3x2pt cosmology \citep[][]{des17}.}. Then, to include projection effects and assign $\lob$ we proceed as in \citetalias{costanzi19}, and compute for each halo:
\begin{equation}
\label{eq:prj}
\lob_i = \ltr_i + \Delta^{\rm prj}_i= \ltr_i + \sum_{j\neq i} \ltr_j f^A_{ij} w_z(z_j,z_i)  \, ,
\end{equation}
where the sum runs over all the halos in projection with $\ltr_j<\ltr_i$, $f^A_{ij}$ is the fraction of area of the $j$-th object in projection which overlaps with the area of the $i$-th target halo, and $w_z(z_i,z_j)$ is the redshift-dependent weight which accounts for the redshift distance between $i$ and $j$ (see equation \ref{eqn:w_z}). 
As mentioned above, this assumes that the galaxies are uniformly distributed inside the cluster radius; while a crude approximation, we have explicitly verified that distributing the galaxies according to a NFW radial profile does not significantly impact the results.
As we neglect the photo-z error on cluster redshift, we set $\zob$ of our mock catalog equal to the true halo redshift. Figure \ref{fig:OMR} shows the scaling relations between mass, $\ltr$ and $\lob$ derived from the mock catalog in the redshift range $0.2<z<0.65$. The effect of projected structures on the observed richness can be clearly seen in the various panels.

To generate mock projected density profiles, we create Monte Carlo realizations of the projected NFW surface density profile by sampling dark matter particle positions around each halo out to a projected radius of $30 \hMpc$.
To ensure well-resolved profiles down to $0.1 \hMpc$ we consider a particle mass of $m_p=10^{10} \hMsun$. Then, for each system, we compute the mean projected density, in 20 equally log-spaced annuli between $0.1$ and $30 \hMpc$, by considering all the particles in a cylinder of depth $\pm 50 \hMpc$ centered on the target halo. The cylinder depth has been chosen to ensure a proper measurement of the contribution from correlated structures up to a projected distance of $\sim 10 \hMpc$ from the cluster center \citep[see e.g.][]{Sunayama20,Wu2022,Zhang23}. We explicitly tested that increasing the cylinder depth to $\pm 100 \hMpc$ does not affect our results. To facilitate comparison with our model predictions, we compute for each cluster, along with the projected density profile, the contribution to the profile due solely to halos in projection along the line of sight, i.e. $\Sigma^{\rm prj}(R)$.

\begin{figure}
    \includegraphics[width=0.45 \textwidth]{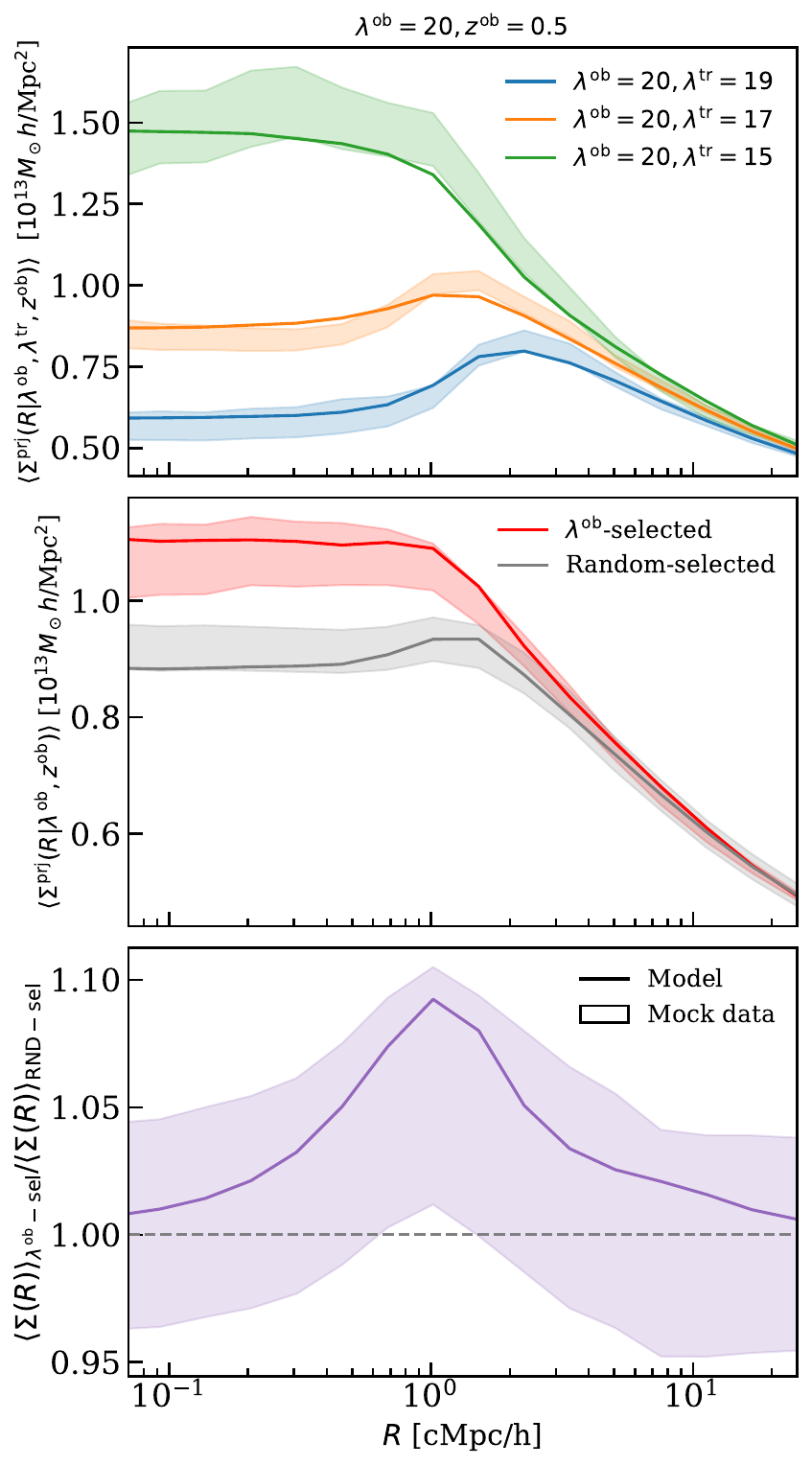}
\caption{Comparison between the model predictions of the density profiles (\textit{solid} lines) and the corresponding measurements from mock data (\textit{shaded bands}) for $\lob=20$ and $\zob=0.5$ clusters. The band width represents the $1\sigma$ uncertainty of the measurements. \textit{Upper} panel: Two-halo component of the projected density profile for different $\ltr$ values. Larger (lower) $\Delta^{\rm prj}=\lob-\ltr$ values correspond to denser (emptier) lines of sight. \textit{Middle} panel: Comparison of the $\langle \Sigma^{\rm prj}\rangle$ profile for a richness-selected sample and a random-selected sample with the same mass distribution. \textit{Lower} panel: Ratio of the richness-selected and random-selected total density profiles. }
\label{fig:Sig_prj}    
\end{figure}
%
\begin{figure*}
\begin{center}
    \includegraphics[width=0.95 \textwidth]{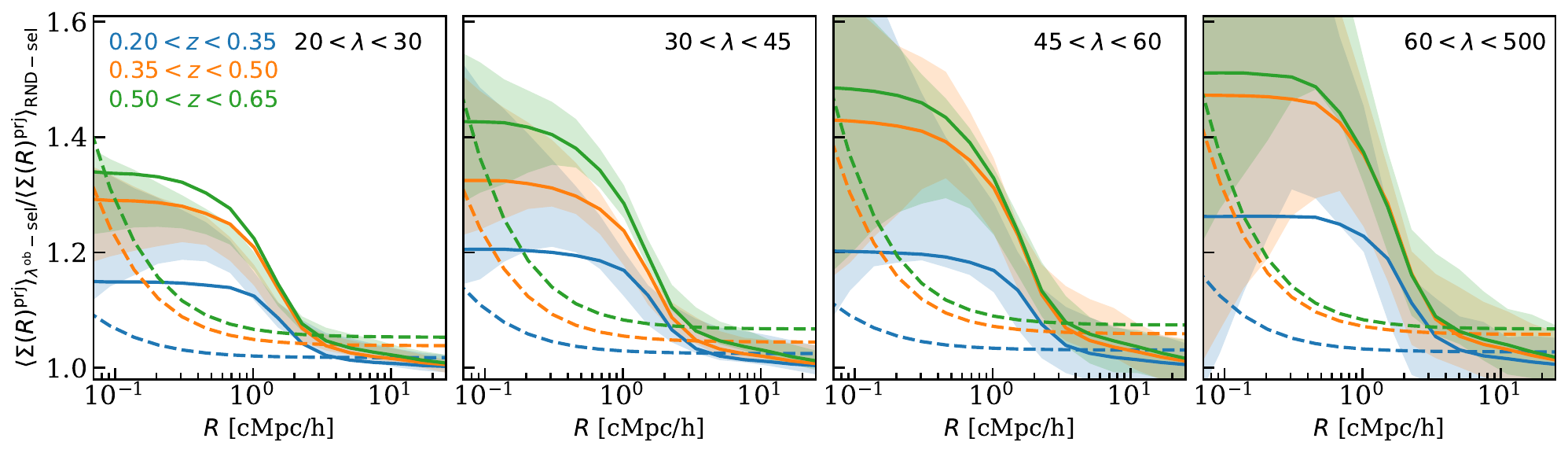}
\end{center}
\caption{Richness and redshift-evolution of the two-halo component of the density profile, $\langle \Sigma^{\rm prj} \rangle$, relative to a randomly selected sample with the same mass and redshift distribution (color-coded \textit{solid} lines). For comparison, the \textit{shaded bands} bracket the $1\sigma$ statistical uncertainty of the same quantity measured from the mock catalog. Also shown with \textit{dashed} lines is the optical cluster bias relative to an unbiased sample, i.e. the ratio $\blob(R)/b_{\rm eff}$.}
\label{fig:sel_bias_SigPRJ_profile}    
\end{figure*}
\section{Results}
\label{sec:res} 

We test our model against the mock catalog considering the same selection criteria  -- $\lob>20$ and $\zob \in [0.2,0.65]$ -- and binning scheme -- $\Delta \lob = [20,30,45,60,\infty]$; $\Delta \zob =[0.20,0.35,0.50,0.65]$ -- as those applied to the DES Y1 and Y3 cluster cosmology catalogs \citep[][]{desy1cl,desy3cl}. 

We start by assessing the consistency of our model for the projected component of the density profile,$\langle \Sigma^{\rm prj} \rangle$,  with mock data. For illustrative purposes, we show in figure \ref{fig:Sig_prj} results for $\lob=20$, $\zob=0.5$ -- roughly the mean richness and redshift of the cosmological sample -- although the same results hold for all the richnesses and redshifts considered in this work. The upper panel compares model predictions and mock measurements of $\langle \Sigma^{\rm prj}(R|\lob,\ltr,\zob) \rangle$ for different values of $\ltr$. The model predictions (\textit{solid} lines) are obtained by substituting $\blob(\lob,\zob,\theta)$ with $\blob(\lob,\ltr,\zob,\theta)$ in equation \ref{eqn:sigprj}. The model is consistent with the synthetic data and correctly reproduces the dependence of the cluster environment on $\dprj$. The \textit{middle} panel shows the 2-halo component of the projected density profile for a richness-selected sample and a random sample sharing the same mass distribution. The model prediction for the latter is obtained by replacing $\blob(\lob,\ltr,z,\theta)$ with $b_{\rm eff}(\ltr,z)$ in equation \ref{eqn:b_sel}. The $\blob$ model proposed here successfully reproduce the typical denser environment along the line of sight of optically selected clusters. Finally, in the \textit{lower} panel is shown the ratio of the projected density profiles derived from a richness-selected and random-selected sample. The optical density profile bias peaks around $\sim 1 \hMpc$, corresponding to the transition scale between the one-halo and two-halo terms. Its bell-shaped form is consistent with findings from similar studies in the literature \citep[see, e.g.,][]{Sunayama20, Wu2022, Salcedo2024}, and can be well described by a double power law with a smooth transition (see Appendix \ref{app:bsel_approx}). Given the computational cost of the model, the latter provides a useful approximation to perform cosmological inference with optical galaxy clusters.

Secondly, we study the evolution of the optical cluster bias and the two-halo component of the density profile with richness and redshift. The panels in figure \ref{fig:sel_bias_SigPRJ_profile} show the evolution of these two quantities relative to a random-selected sample with the same mass and redshift distribution as the $(\lob, \zob)$ bin under consideration. Both $\blob$ (\textit{dashed} lines) and $\langle \Sigma^{\rm prj} \rangle$ (\textit{solid} lines) exhibit a mild evolution with richness and redshift.  The redshift trend can be mainly understood in terms of the evolution of the photo-z uncertainty, which increases with distance from the observer, whereas the richness dependence reflects the denser environments of richer, and therefore on average more massive, systems.

\begin{figure*}
\begin{center}
    \includegraphics[width=0.85 \textwidth]{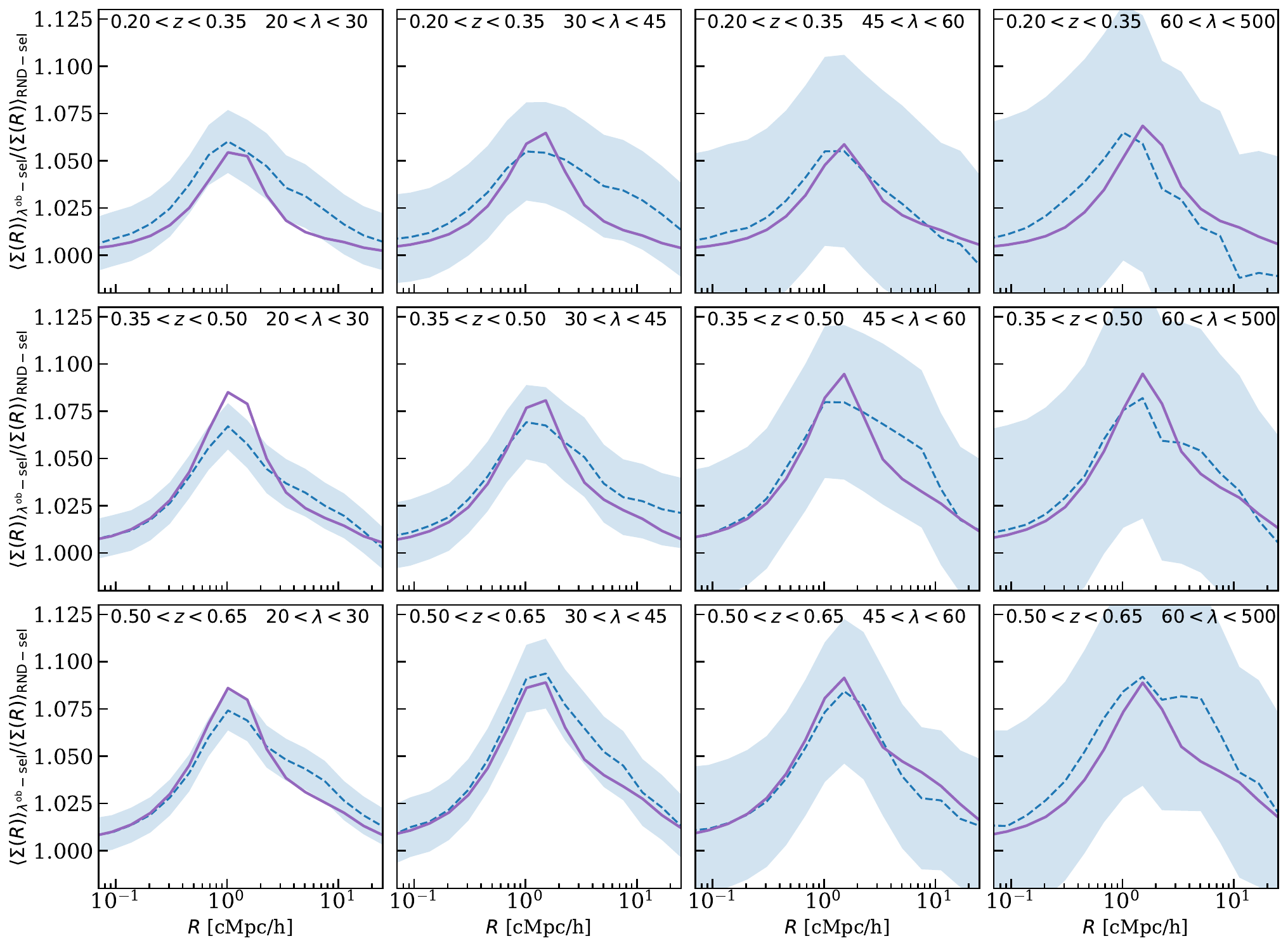}
\end{center}
\caption{Optical selection bias on cluster lensing profiles for different richness and redshift bins (see panels legend). The \textit{purple solid} lines correspond to the model predictions, while the \textit{dashed blue} lines are the mean measurements from the mock catalog along with the $1\sigma$ statistical uncertainty (\textit{shaded} bands).}
\label{fig:sel_bias_profile}    
\end{figure*}

Finally, we look at the evolution of the overall bias on the lensing profile as a function of richness and redshift. Figure \ref{fig:sel_bias_profile} shows the ratio of the total density profiles derived from optically-selected and random-selected samples in different bins. Again, the model (\textit{purple} solid lines) reproduces the mock data within the statistical uncertainty (\textit{dashed} lines and \textit{shaded} bands). In this case, there is no clear trend with richness and redshift as the evolution of the two-halo term is compensated by the evolution of the one-halo term to the total density profile. These results can be qualitatively compared to those presented in \citet[][]{Sunayama20}, \citet{Wu2022}, \citet{Zeng2023} or \citet{Salcedo2024}; however, the exact magnitude and evolution of the bias depend on the specifics of the survey and the cluster finder under consideration.

\section{Summary and conclusions}
\label{sec:conc}

In this paper, we presented an analytical framework to predict the selection bias affecting the projected density profiles of photometric cluster catalogs due to projection effects. In particular, we modeled the scale-dependent clustering bias of optically selected clusters, $\blob$, as a function of the richness contamination induced by line-of-sight structures. This relation was then used to compute the mean contribution of correlated structures to a cluster projected density profile by integrating the off-axis density profiles of projected halos.

As a case study, we considered a DES Y3-like photometric cluster sample and generated a synthetic cluster catalog that self-consistently accounts for projection effects on both richness and projected density profiles. Motivated by the analysis of the cluster correlation function of richness-selected and random-selected mock samples, we introduced a sigmoid parameterization to describe $\blob$ as a function of projected angular distance from the cluster center. We showed that this model successfully reproduces the dependence of the cluster bias on the magnitude of projection effects, $\dprj = \lob - \ltr$. The distortion induced by selection effects is strongest at small projected separations but persists out to scales of $R \simeq 10 \hMpc$. We further found that the large-scale value of the optical cluster bias can be well approximated by a linear function of $\delta^{\rm prj}$, the excess of richness bias induced by projection effects relative to a random sample with the same mass distribution.

From the model for $\blob$, we predict the mean number of halos projected along the line of sight to a target cluster with given $\lob$ and $\zob$, and thus estimate their contribution to the projected density profile. Consistent with their clustering properties, the projected component of the profile, $\langle \Sigma^{\rm prj} \rangle$, depends on the magnitude of $\dprj$, implying that, on average, $\lob$-selected clusters reside in denser environments than random-selected samples with the same mass distribution. All model predictions accurately reproduce the corresponding measurements from the synthetic DES Y3 cluster catalog, including their scale dependence up to $R \sim 10 \hMpc$ and their evolution with richness and redshift. In particular, the model captures the characteristic bell-shaped bias observed in the total density profiles of optically selected clusters.

The simulation tests performed here were designed to validate the core analytical equations governing the small-scale selection bias (Eq. \ref{eqn:bsel0}) and the projected component of the density profile (Eq. \ref{eqn:sigprj}). In general, the magnitude of $\blob$ and of the projected density bias predicted by the model depends on the specifics of the cluster finder and survey, primarily through the projection kernel $w_z(z,\zob)$ and the conditional probability distribution $P(\lob|\ltr)$. Both quantities can be empirically calibrated using simulation-based approaches or multi-wavelength data. The framework is also readily generalizable to alternative optical mass proxies, such as the probability-weighted stellar mass $\mu_{\star}$ \citep{Esteves2025} or the number of galaxies $N_{\rm gals}$ used by algorithms like WaZP \citep{Aguena2021}, provided that the definition of the projection weight $\rho_{\rm prj}$ (equation \ref{eqn:weight}) is adapted accordingly.

Despite its simplicity—for example, the present analysis neglects cluster miscentering as well as selection effects related to halo orientation and triaxiality—this work provides a physically motivated framework to interpret and predict selection biases in the weak-lensing profiles of photometric cluster catalogs.
In cosmological applications, it is important to account for systematic uncertainties associated with the modeling of projection effects, including the choice of projection kernel and geometric factor $f_A$, the calibration of $P(\lob|\ltr)$, and the assumed functional form of the optically selected cluster bias $\blob$. Moreover, given the large number of nested integrals required to compute the model predictions, direct implementation in cosmological parameter-inference pipelines may be computationally expensive. For such applications, we provide in Appendix \ref{app:bsel_approx} a double power-law fitting function that approximates the bias in the projected density profile. The parameters of this fitting function can be assigned priors calibrated from our model, enabling efficient marginalization over systematic uncertainties in the projection modeling, as well as over HOD and cosmological parameters in cosmological analyses.

Finally, although this framework was developed to describe projected density profiles of optically selected clusters on scales $R \lesssim 10 \hMpc$, it could in principle be extended to model their large-scale clustering properties. We caution, however, that the accuracy of the large-scale cluster bias prescription, $\blob^\infty$, has not been tested across different cosmologies and may require further refinement for precision clustering analyses.


\section*{Acknowledgments}

MC and MA are supported by the PRIN 2022 project EMC2 - Euclid Mission Cluster Cosmology: unlock the full cosmological
utility of the Euclid photometric cluster catalog (code no. J53D23001620006). 
HW is supported by the US DOE Award DE-SC0010129 and the NSF Award AST-2509910.
JHE is supported by Harvard University and by the US DOE award DE-SC0007881.
We gratefully acknowledge the use of the \texttt{NumPy} and \texttt{SciPy} Python libraries \citep{harris2020array,virtanen2020scipy} for numerical computations and scientific routines, and of \texttt{Matplotlib} \citep{hunter2007matplotlib} for producing the figures.
The linear and non-linear matter power spectra were computed using the \texttt{CAMB} Boltzmann code \citep{Lewis:2002ah}. Part of the analysis was performed using resources of the computing center of INAF-Osservatorio Astronomico di Trieste \citep{taffoni2020,bertocco2019}.




\appendix

\section{Analytical model for $\blob$}
\label{app:bsel}
\begin{figure}
    \includegraphics[width=0.48 \textwidth]{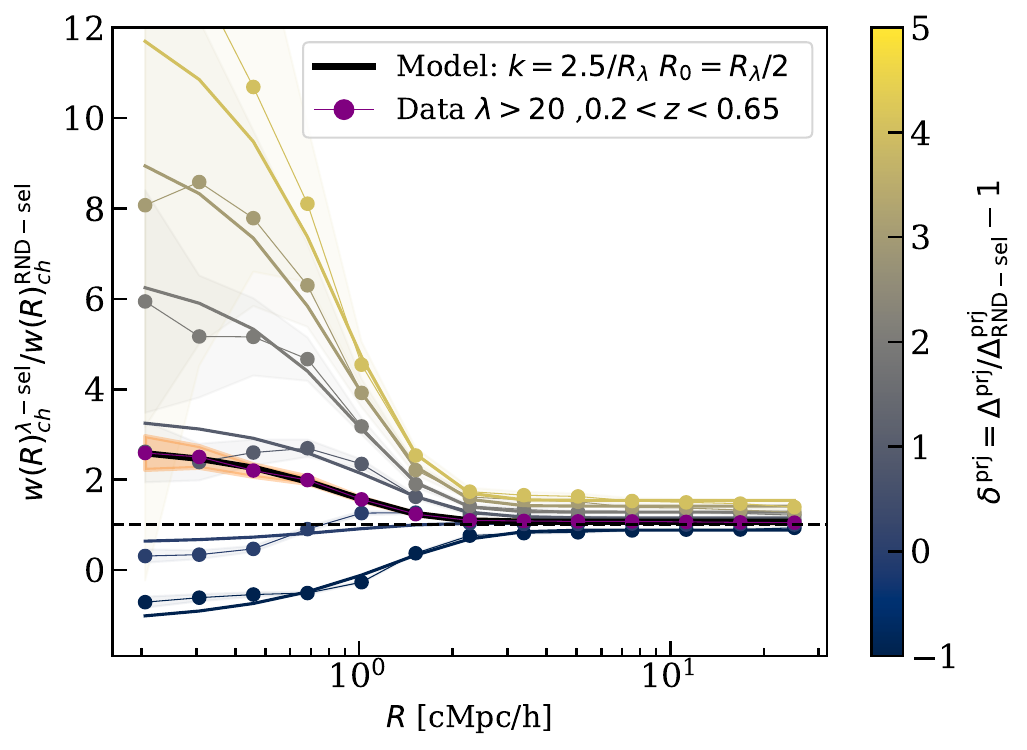}
\caption{Ratio of cluster–halo correlation functions of richness-selected and randomly-selected sample with the same mass distribution. Dots represent measurements from the mock catalog while the solid lines corresponds to the best-fit sigmoid model fixing $k=2.5/R_\lambda$ and $R_0=R_\lambda/2$ (equation~\ref{eqn:b_sel_theta}). Color-coded lines and dots correspond to sub-samples of the catalog with different value of $\dprj=\lob-\ltr$ normalized to the $\dprj$ value of a mass-selected sample. The \textit{purple} dots and \textit{black} line correspond to the measurements and best-fit model for the whole cosmological sample.}
\label{fig:wch_wmh}    
\end{figure}

To derive an empirical model for the optical cluster bias, we consider the ratio of the projected cluster-halo correlation functions derived from our mock catalog for a richness-selected sample and a random-selected sample having the same mass distribution. This ratio corresponds to the ratio of the cluster biases of the two samples, i.e. $\blob / b_{\rm eff}$, where $b_{\rm eff}(\lob,z) \propto \int dM P(\lob|M,z) b(M,z) n(M,z) $ \citep[e.g.][]{desy3cl_method}. Figure \ref{fig:wch_wmh} shows these measurements along with the best-fit sigmoid model proposed in this work (equation \ref{eqn:b_sel_theta}), fixing the shape parameters to $k=2.5/\theta_\lambda$ and $\theta_0=\theta_\lambda/2$.
We note that \citet{desy3cl_method} found a qualitatively similar functional form using a mock redMaPPer catalog obtained by running the cluster finder directly on synthetic galaxy catalogs and therefore not relying on our specific assumptions for $w_z$ and $f_A$.
To highlight the dependence of $\blob$ on the magnitude of projection effects, we split the catalog according to $\delta^{\rm prj} = \dprj/\dprj_{\rm RND-sel} -1$, which quantifies the excess of projection effects compared to a random-selected sample (color-coded dots and lines in figure; see equation \ref{eqn:bsel_infty}). These results show that larger (smaller) $\delta^{\rm prj}$ values correspond to denser (emptier) environments around the cluster, and that this effect persists up to large projected distances \citep[$R \gtrsim 10 \hMpc$; see also][]{Sunayama20,Zeng2023}.  
The sigmoid model provides a good fit to the selection bias measured from the sub-samples, as well as to that inferred from the full cosmological sample (\textit{purple} dots and \textit{black} line). 

\begin{figure}
    \includegraphics[width=0.48 \textwidth]{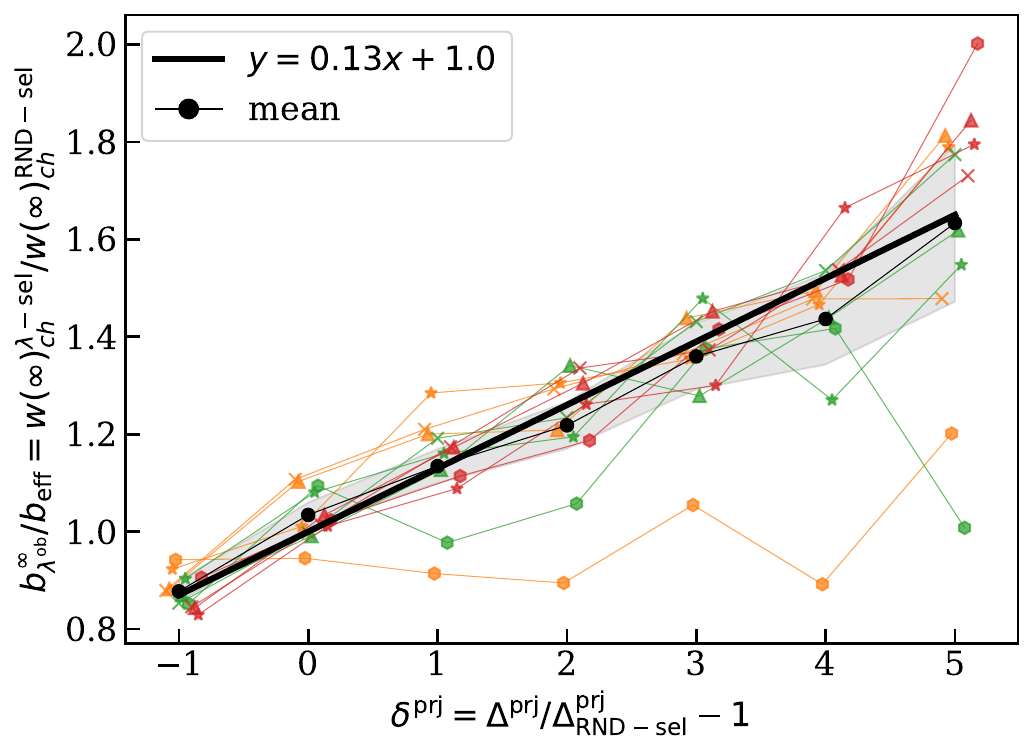}
\caption{Large-scale selection bias, $\blob^\infty / b_{\rm eff}$, as a function of $\dprj=\lob-\ltr$ normalized to the $\dprj$ value of a random-selected sample with the same mass distribution. The data points correspond to measurements in different redshift (\textit{yellow}: $0.2<z<0.35$; \textit{green}: $0.35<z<0.50$; \textit{red}: $0.50<z<0.65$) and richness bins (stars: $20<\lob<30$; crosses: $30<\lob<45$; triangles: $45<\lob<60$; dots: $60<\lob<\infty$) bins. The \textit{black} dots with the gray shaded band and solid line represent the mean measurements (with $1\sigma$ uncertainty) over richness and redshift bins, and the best-fit linear relation, respectively. }
\label{fig:b_sel_infty}    
\end{figure}

To close the system of equations and obtain a predictive model for $\blob$, we must constrain one additional degree of freedom. To this end, we model the boost of the large-scale bias relative to that of a random-selected sample, $\blob^\infty/b_{\rm eff}$, as a function of $\delta^{\rm prj}$. As shown in figure \ref{fig:b_sel_infty}, this quantity exhibits no clear dependence on richness or redshift (different colors and markers), and its mean relation (\textit{black} dots) is well approximated by a linear function: $\blob^\infty/b_{\rm eff} = 1 + 0.13 \delta^{\rm prj}$.

As shown in the main text, this model is sufficiently accurate to reproduce mock data measurements of the projected density profiles within their statistical uncertainties.

\begin{figure}
    \includegraphics[width=0.45 \textwidth]{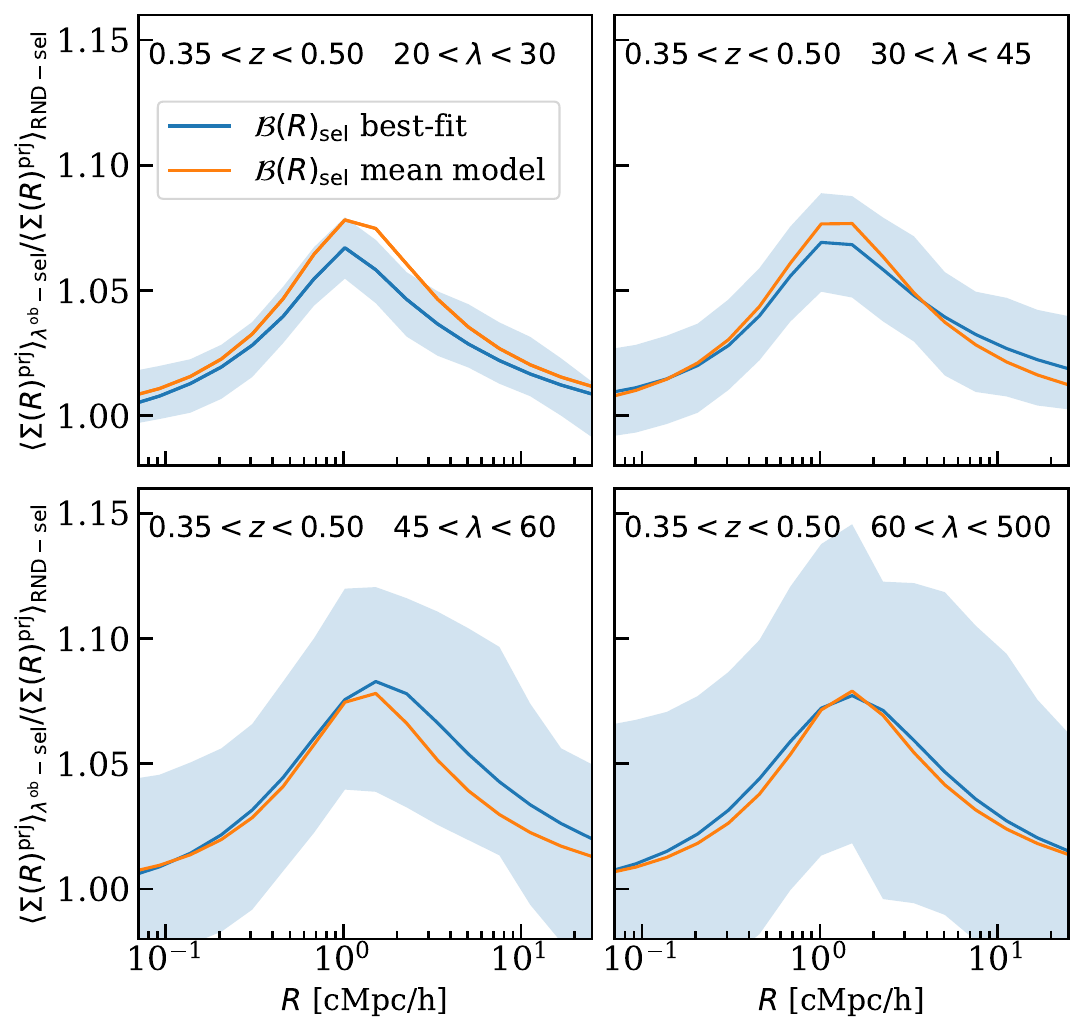}
\caption{Comparison of the analytical fitting function $ \mathcal{B}(R)_{\rm sel}$ (\textit{solid} lines) with mock data (\textit{shaded} bands) in four richness bins at $0.35<z<0.50$. The \textit{blue} lines correspond to the best-fit model derived in each richness bins, while the \textit{orange} line is the mean best-fit model derived considering the whole catalog. }
\label{fig:sel_bias_Sig_profile_approx}    
\end{figure}

\section{Fraction overlapping area}
\label{app:a}
Given a model for the normalized projected galaxy radial distribution, $\Sigma_{\rm glx}(R)$, and assuming spherical symmetry, the fraction of member galaxies of a halo in projection that falls within the area of a target cluster is given by:
\begin{eqnarray}
    f_A(\theta,\theta_{\lambda},\theta_{\rm h}) &=& \int_0^{\theta_\lambda} d\vartheta \vartheta \int_{\varphi_{\rm min}(R)}^{\varphi_{\rm max}(R)} d\varphi \Sigma_{\rm glx}\left( R_h \right) \\
    R_h &=& \frac{\sqrt{\vartheta ^2 + \theta^2-2\vartheta\theta\cos(\varphi)}}{D_A(z)} \nonumber \\
    \varphi_{\rm max}(R) &=& - \varphi_{\rm min}(R) = \arccos\!\left( 
    \frac{\vartheta^2 + \theta_\lambda^2 - \theta_h^2}
         {2\vartheta\theta_\lambda} \right). \nonumber
\end{eqnarray}
Here, $\theta$ is the angular separation between the two centers, while $\theta_\lambda=R_\lambda/D_A(\zob)$ and $\theta_h=R_h/D_A(z_h)$ denote the angular sizes of the target cluster and the halo in projection, respectively. Their projected physical sizes are computed from the observed cluster and true halo richness using the empirical calibrated \redmapper\ radius $R_\lambda = (\lambda/100)^{0.2} \hMpc$. 
Assuming that the galaxies are uniformly distributed within the halo, the fraction simplifies to:
\begin{equation}
f_A(\theta,\theta_{\lambda},\theta_{\rm h}) = \frac{A_\text{ov}(\theta,\theta_{\lambda},\theta_{\rm h})}{\pi \min(\theta_\lambda^2, \theta_h^2)}
\end{equation}
where $A_\text{ov}$ is the overlapping area, calculated as:
\begin{equation}   
A_\text{ov} =
\begin{cases}
0, & \text{if } \theta \geq \theta_\lambda + \theta_h, \\
\pi \min(\theta_\lambda^2, \theta_h^2), & \text{if } \theta \leq |\theta_\lambda - \theta_h|, \\
A_\cap(\theta,\theta_\lambda,\theta_h), & \text{otherwise.}
\end{cases}
\end{equation}
with
\begin{multline}
    A_\cap(\theta,\theta_\lambda,\theta_h) = \theta_\lambda^2 \cos^{-1}\left(\frac{\theta^2 + \theta_\lambda^2 - \theta_h^2}{2 \theta \theta_\lambda}\right) \\
+ \theta_h^2 \cos^{-1}\left(\frac{\theta^2 + \theta_h^2 - \theta_\lambda^2}{2 \theta \theta_h}\right)  
 - \frac{1}{2}  [(-\theta + \theta_\lambda + \theta_h) \\
 (\theta + \theta_\lambda - \theta_h)(\theta - \theta_\lambda + \theta_h)(\theta + \theta_\lambda + \theta_h)]^{1/2}
\end{multline}

\section{Analytical fitting function for the selection bias on projected density profile}
\label{app:bsel_approx}
For practical purposes, we provide here an analytical expression to approximate the optical selection bias on the projected density profile. The bias can be described by a double power-law with smooth transition, namely:
\begin{equation}
    \label{eqn:bsel_sigma}
    \mathcal{B}_{\rm sel}(R) = A \left( \frac{R}{R_0} \right)^\alpha  \left[ 1+ \left(\frac{R}{R_0}\right)^{\gamma} \right]^{(\beta - \alpha)/\gamma} + 1
\end{equation}
where $A$ sets the amplitude, $\alpha$ controls the slope at small radii, $\beta$ the slope in the outer region, $R_0$ defines the transition scale, and $\gamma$ determines the smoothness of the transition.
Setting the transition scale $R_0$ to the comoving cluster radius $R_\lambda(1+z)$, the best-fit parameters derived considering the full mock catalog are: $A=0.10$, $\alpha=0.92$, $\beta=-0.53$, $\gamma=4.1$ (see figure \ref{fig:sel_bias_Sig_profile_approx}).
For completeness, we note that the same functional can also approximate the bias on the excess surface density profile, $\Delta \Sigma$, a quantity more directly related to the observable weak-lensing shear profile.

\end{document}